\documentclass[twocolumn, secnumarabic, aps, prl, superscriptaddress, amssymb, amsmath, nobibnotes, showpacs]{revtex4}
\usepackage{graphicx}

\begin{document}
\title{Facilitated diffusion of DNA-binding proteins}

\author{Konstantin V.\ Klenin}
\affiliation{Division of Biophysics of Macromolecules,
German Cancer Research Center, D-69120 Heidelberg, Germany}
\author{Holger Merlitz}
\email{merlitz@gmx.de}
\affiliation{Department of Physics, Xiamen University,
Xiamen 361005, P.R.\ China}
\author{J\"org Langowski}
\email{jl@dkfz.de}
\affiliation{Division of Biophysics of Macromolecules,
German Cancer Research Center, D-69120 Heidelberg, Germany}
\author{Chen-Xu Wu}
\affiliation{Department of Physics, Xiamen University,
Xiamen 361005, P.R.\ China}

\date{\today}
\begin{abstract}
The diffusion-controlled limit of reaction times for
site-specific DNA-binding proteins is derived
from first principles.
We follow the generally accepted concept that a protein
propagates via two competitive modes, a
three-dimensional diffusion in space and
a one-dimensional sliding along the DNA.
However, our theoretical treatment of the problem is new.
The accuracy of our analytical model is verified
by numerical simulations. The results confirm that
the unspecific binding of protein to DNA, combined with
sliding, is capable to reduce the reaction
times significantly.
\end{abstract}
\maketitle

\textbf{Introduction.} The understanding of diffusion controlled chemical
reactions has become an indispensable ingredient of
present days technological development. The optimization of catalysts,
fuel cells, improved batteries using electrodes with nano-structured
surfaces or the function of semi-conductive devices are just a
few of countless examples where diffusive processes,
often in crowded or fractal environments, are involved
to define the most important system parameters. For any living
organism, diffusion plays the central role in  biochemical and
-physical reactions that keep the system alive~\cite{riggs70,richter74}:
The transport of molecules through cell membranes, of
ions passing the synaptic gap or drugs on the way to
their protein receptors are predominantly diffusive
processes. Further more, essentially all of the biological
functions of DNA are performed by proteins that interact
with specific DNA sequences~\cite{berg85, ptashne01},
and these reactions are diffusion-controlled.

However, it has been realized that some proteins
can find their specific target sites on DNA
much more rapidly than is ``allowed'' by the diffusion
limit~\cite{riggs70, berg81, halford04}.
It is therefore generally
accepted that some kind of facilitated diffusion must
take place in these cases.
Several mechanisms, differing in details, have been proposed for it.
All of them essentially involve
two steps. First, the protein binds to a random non-specific
DNA site. Second, it diffuses (slides) along the DNA chain.
These two steps may be reiterated
many times before the protein actually finds the target, since the sliding is
occasionally interrupted by dissociation.

Berg et al.\
have provided a thorough (but somewhat sophisticated) theory
that allows an estimation of the resulting reaction rates~\cite{berg81}.
Recently, Halford and Marko have presented
a comprehensive review on this subject and proposed
a remarkably simple semiquantitative approach
that explicitly contains the mean sliding length
as a parameter of the theory~\cite{halford04}.

In the present
work we suggest an alternative view on the problem
starting from first principles.
Our theory leads to a formula that is similar in form to that of
Halford and Marko, apart from numerical factors.
In particular, we give a new interpretation of the sliding length,
which makes it possible to relate this quantity
to experimentally accessible parameters.

\textbf{Theory.} To estimate the mean time $\tau$ required
for a protein to find its target,
we consider a single DNA chain in a large volume $V$. At time $t=0$, the
protein molecule is somewhere outside the DNA coil. We introduce the
`reaction coordinate' $r$ as the distance between the center of the
protein and the center of the target, which is assumed to be
presented in one copy. When $r$ is large, the only
transport mechanism is the 3-dimensional (3d) diffusion in space. On the
contrary, at small $r$, the 1-dimensional (1d) diffusion along the
DNA chain is more efficient.

Let us define the efficiency of a transport mechanism in more
strict terms. Let $\tau(r-dr,r)$ be the mean time of the first
arrival of the protein at the distance $(r-dr)$ from the target,
provided it starts
from the distance $r$. In the simple cases, when the diffusion
of a particle can be fully characterized by a single
coordinate, this time is given by the equation~\cite{szabo80, klenin04}
\begin{equation} \label{eq05}
d\tau \equiv \tau(r-dr,r) = \frac{Z(r)}{D\, \rho(r)}\, dr\;,
\end{equation}
where $D$ is the
diffusion coefficient, $\rho(r)$ the equilibrium distribution
function of the particle along the reaction coordinate (not necessary
normalized), and $Z(r)$ the local normalizing factor
\begin{equation}  \label{eq10}
Z(r) = \int_r^\infty \rho(r')\,dr'\;.
\end{equation}
Note that the quantity $1/d\tau$ is the average frequency of transitions
$r \rightarrow r-dr$ in the `reduced' system with a reflecting boundary
at the position $r-dr$ (so that the smaller distances from the target are forbidden).
The quantity
\begin{equation}  \label{eq15}
v \equiv \frac{dr}{d\tau} = \frac{D\, \rho(r)}{Z(r)}
\end{equation}
has the dimension of velocity and can be regarded as a
measure for the efficiency of a transport process.

For 3d-diffusion inside the volume \textit{V}, we have $\rho(r) = 4\pi\, r^2 c$, where $c$ is
the protein concentration and the factor $4\pi$ is chosen to provide a convenient normalization for a system containing only one protein molecule: $Z(0) = Vc = 1$.
Hence, for sufficiently small $r$, when $Z(r) \approx Z(0) = 1$, the transport efficiency is
\begin{equation}  \label{eq20}
v_{\text{3d}}(r) = 4 \pi D_{\text{3d}} r^2 c\;.
\end{equation}
In the case
of a 1d-diffusion along the DNA chain we have $\rho(r) = 2\sigma$,
with $\sigma$ being the linear density of a non-specifically bound
protein. The factor 2 accounts for the fact that the target can
be reached from two opposite directions.  We assume, again,
that the distance $r$ is sufficiently
small, so that the DNA axis can be considered as a straight line.
Thus, the efficiency of the 1d-diffusive transport near the target
is given by
\begin{equation}  \label{eq25}
v_{\text{1d}} = 2 D_{\text{1d}} \sigma\;.
\end{equation}
Our main assumption is that, during the combined diffusion process,
the probability of the (non-specifically) bound state
is close to its equilibrium value for each given value of $r$. Then the
frequencies $1/d\tau_{\text{3d}}$ and $1/d\tau_{\text{1d}}$ are
additive, and so are the
efficiencies of the two transport mechanisms given by
Eqs.~(\ref{eq20}) and (\ref{eq25}). Hence, the mean time of the first arrival
at the target of radius $a$ can be found as
\begin{equation} \label{eq30}
\tau = \int_a^\infty \frac{dr}{v_{\text{3d}} +  v_{\text{1d}}}\;.
\end{equation}
The main contribution to this integral is made by the distances
close to $a$. For that reason, the upper limit of integration is set
to infinity. Before evaluation of Eq.~(\ref{eq30}), we note that
\begin{equation} \label{eq31}
1 = Z(0) = Vc + L\sigma\;,
\end{equation}
where $V$ is the volume and $L$ is the DNA length.
The meaning of this equation is that the system contains only one protein
molecule.
Substituting Eqs.~(\ref{eq20}) and (\ref{eq25}) into
Eq.~(\ref{eq30}) and taking into account Eq.~(\ref{eq31}),
we get, finally,
\begin{equation} \label{eq32}
\tau = \left( \frac{V}{8D_{\text{3d}}\,\xi} +
    \frac{\pi\,L\, \xi}{4 D_{\text{1d}}} \right) \left[
    1 - \frac{2}{\pi} \arctan \left(\frac{a}{\xi}\right)\right]\;.
\end{equation}
Here, we have introduced a new parameter
\begin{equation} \label{eq35}
\xi = \sqrt{\frac{D_{\text{1d}}\, K}{2\pi\, D_{\text{3d}}}}\;,
\end{equation}
with $K=\sigma/c$ being the equilibrium constant of non-specific binding.
It is easy to verify that $\xi$ is just the distance, where the efficiencies
of the two transport mechanisms [Eqs.~(\ref{eq20}) and (\ref{eq25})]
become equal to each other.

\textbf{Numerical model.} In what follows we present numerical simulations
to test the accuracy of our analytical result for the reaction time
given by Eqs.~(\ref{eq32}) and (\ref{eq35}). In order to approximate
the real biological situation,
the DNA was modeled by a chain of $N$ straight segments of equal length $l_0$.
Its mechanical stiffness was defined by the bending energy
associated with each chain joint:
\begin{equation} \label{eq56}
E_b = k_B T \, \alpha \, \theta^2\;,
\end{equation}
where $k_B T$ is the Boltzmann factor, $\alpha$ the dimensionless
stiffness parameter, and $\theta$ the bending angle. The numerical
value of $\alpha$ defines the persistence length, i.e.\ the 
``stiffness'' of the chain~\cite{klenin98}.
The excluded volume effect was taken into account by introducing
the effective DNA diameter, $d_{\text{eff}}$. The conformations of the chain,
with the distances between non-adjacent segments smaller than $d_{\text{eff}}$,
were forbidden.
The target of specific binding was assumed to lie exactly in the
middle of the DNA.
The whole chain was packed in a spherical volume (cell) of radius $R$
in such a way that the target occupied the central position.

In order to achieve a close packing of the chain inside the
cell, we first generated a relaxed conformation of the free chain
by the standard Metropolis Monte-Carlo (MC) method.
For further compression, we defined the
center-norm (c-norm) as the maximum distance from the target
(the middle point) to the other parts of the chain.
Then, the MC procedure was continued, but a MC step was rejected if the
c-norm was exceeding 105\% of the lowest value registered so
far. The procedure was stopped when the desired degree of compaction
was obtained.

The protein was modeled as a random walker within the cell
with reflecting boundaries.
During one step in the free 3d-mode, it was displaced by the
distance $\varepsilon_{\text{3d}}$ in a random direction.
Once the walker approached the chain closer than a certain capture
radius $r_{\text{c}}$, it was placed to the nearest point on the chain
and its movement mode was changed to the 1d-sliding along
the chain contour. In this mode, the step represented a
displacement by the distance $\varepsilon_{\text{1d}}$ performed
with an equal probability in either direction.
The ends of the chain were reflective.
After each 1d-step (and immediately after the capture) the walker
could jump off the chain by the distance $r_{\text{c}}$ and reenter the 3d-mode.
This operation was carried out with the kick-off probability $p$.

A simulation cycle started with the walker at the periphery
of the cell and ended when the walker came within
the distance $a$ to the target. During all simulation cycles the chain
conformation remained fixed. 

Below in this paper, one step is chosen as the unit of time and
one persistence length of the DNA chain (50~nm) as the unit of distance.
The following values of parameters were used. The length of one segment
was chosen as $l_0 = 0.2$, so that one persistence length was partitioned into
5 segments. The corresponding value of the stiffness parameter was
$\alpha = 2.403$~\cite{klenin98}.
The effective chain diameter was $d_{\text{eff}} = 0.12$, the
capture radius $r_{\text{c}} = d_{\text{eff}}/2$, and the radius of the active
site was $a = 0.08$. The diffusion coefficients are defined as
$D_{\text{3d}} = \varepsilon_{\text{3d}}^2/6$ and
$D_{\text{1d}} = \varepsilon_{\text{1d}}^2/2$.
The step-size of the walker was $\varepsilon_{\text{3d}} = 0.04$ and
$\varepsilon_{\text{1d}} = \varepsilon_{\text{3d}}/\sqrt{3}$, yielding
identical diffusion coefficients
$D_{\text{3d}} = D_{\text{1d}} = 8\cdot 10^{-4}/3$.

The radius $R$ of the cell and the DNA length $L$ were varied
in different sets of simulation. For each fixed pair ($R$,$L$),
the kick-off probability was
initially set to $p = 1$ (no 1d-transport, $\xi = 0$)
and subsequently reduced to $p_i \equiv 2^{-i}$, $i = 1, 2, \dots, 11$.
For each parameter set, the simulation cycle was repeated 2000 times.
The equilibrium constant $K$ required for the calculation of the parameter
$\xi$ [Eq.~(\ref{eq35})] has to be determined as the ratio
$V \tau_{\text{1d}} / L \tau_{\text{3d}}$, where
$\tau_{\text{1d}}$ and $\tau_{\text{3d}}$ are the average times the walker
spent in the bound and the free states, respectively.
Note that $\xi$ depends on the choice of the probability $p$,
but not on cell size or chain length, since $\tau_{\text{1d}} \sim L$
and $\tau_{\text{3d}} \sim V$. For each choice of
$p$, the constant $K$ was determined in a special long simulation
run without target for specific binding.

\begin{figure}
\includegraphics[width=1.0\columnwidth]{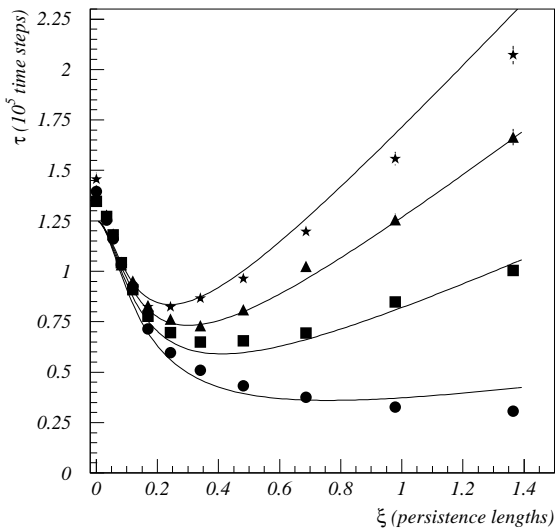}
\caption{Reaction time $\tau$ as a function of the sliding
parameter $\xi$ [Eq.~(\protect\ref{eq35})] at a fixed cell radius
$R = 2$ and chain lengths
$L = 56,\; 40,\; 24,\; 8$ (top to bottom).
The curves are plots of Eq.~(\protect\ref{eq32}).
\label{figR50}}
\end{figure}

\textbf{Results.} In a first set of simulations, chains of various lengths
between $L = 8$ and $L = 56$ were packed into a
cell of radius $R = 2$ and volume $V_0 = 4\pi R^3/3 = 32\pi/3$.
The resulting averaged reaction times $\tau$  are
plotted in Fig.\ \ref{figR50} as a function of the
variable $\xi$ [Eq.~(\ref{eq35})]. The curves are
plots of Eq.~(\ref{eq32}).
It is obvious that the above relation was well able
to reproduce the simulation results on a quantitative
level. This good agreement between theoretical
and computational model indicates that the derivation
of Eq.~(\ref{eq32}), although quite simple, already
contains the essential ingredients of the underlying
transport process. A moderate deviation between simulation
and theory is visible in case of $L = 56$ and large
values of $\xi$. In the discussion we will shortly touch
the limits of the theoretical approach if $\xi$ becomes
very large. With the
present selection of chain-parameters, the results prove
that a 1d-sliding can speed up the reaction time
significantly. If, however, the unspecific binding
becomes too strong, its effect turns into the opposite
and the reaction time is increasing. The most efficient
transport is achieved with a balanced contribution of
both 1d- and 3d-diffusion.

\begin{figure}
\includegraphics[width=1.0\columnwidth]{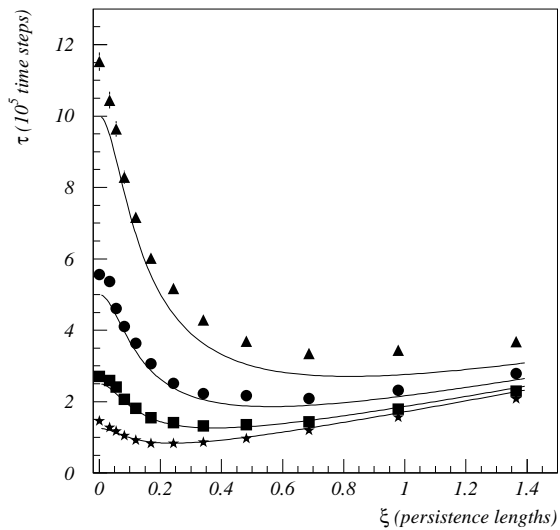}
\caption{Reaction time $\tau$ as a function of the sliding
parameter $\xi$ [Eq.~(\protect\ref{eq35})] at fixed chain
length $L = 56$ and with varying cell volumes
(8x, 4x, 2x and 1x the original volume $V_0 = 32\pi/3$, top to
bottom). The curves are plots of Eq.~(\protect\ref{eq32}).
\label{figL1400}}
\end{figure}

Figure \ref{figL1400} displays
the results of a second set of simulations, where the longest chain
of $L = 56$ was placed into cells with volumes of two, four and eight times
the initial value $V_0 = 32\pi/3$, leading to
systems of rather
sparse chain densities. The plots of Eq.~(\ref{eq32})
are again in good overall agreement with the simulation
results, although a systematic deviation in case of
large cell volumes, i.e.\ at low chain densities, is
visible. The theoretical approach seems to under-predict the
reaction time by up to 10\%. A systematic investigation
of the limits of our approach is part of ongoing research.
For the time being we note that in crowded environments
(of high chain density) Eq.~(\ref{eq32}) appears to be more
accurate than in sparse environments.  

\textbf{Discussion.} Recently, Halford and Marko have proposed
a remarkably simple semiquantitative approach to estimate
the reaction time~\cite{halford04}, yielding the expression
\begin{equation} \label{eq60}
\tau =  \frac{V}{D_{\text{3d}}\,l_{\text{sl}}} +
    \frac{L\, l_{\text{sl}}}{D_{\text{1d}}}
\end{equation}
Following their argumentation, $l_{\text{sl}}$ was interpreted as
the average sliding-length of the protein on the DNA contour.
It is instructive to note that, for $\xi \gg a$,
Eq.~(\ref{eq32}) turns into
\begin{equation} \label{eq65}
\tau = \frac{V}{8D_{\text{3d}}\,\xi} +
    \frac{\pi\,L\, \xi}{4 D_{\text{1d}}} \;,
\end{equation}
which is of identical functional form if we identify $\xi$
with the sliding length of Halford and Marko. With Eq.~(\ref{eq35}) we can now
express $l_{\text{sl}}$ in terms of experimentally
accessible quantities, assigning a physical meaning to
a previously heuristic model parameter. Additionally,
Eq.~(\ref{eq65}) contains the numerical factors which turn
the initially semi-quantitative approach into a model of
quantitative accuracy.

Our results demonstrate (Fig. 2) that crowding decreases the optimum sliding length: the shortest reaction time is reached at lower non-specific binding affinities. In a crowded environment the chance for the protein to bind or re-bind non-specifically is much higher, so that the period of free diffusion is shorter after each kick. In contrast, in sparse environments the chance to hit the target is increased if the protein remains in sliding mode over a rather long distance. 
Increasing the chain density will shift the minimum of $\tau$ to lower values of $\xi$ (Fig. 2), while decreasing the chain length {\it at constant volume} will shift it to higher values (Fig. 1). 
The derivative of Eq.~\ref{eq65} allows an estimate of the optimum sliding
length $\xi_{\text{opt}}$:
\begin{equation} \label{eq66}
\xi_{\text{opt}} = \sqrt{\frac{V\,D_{\text{1d}}}{2\pi\,L\, D_{\text{3d}}}} \;
\end{equation}

Sliding distances have been estimated experimentally to up to 1000 bp for the restriction endonuclease EcoRV in dilute solution from the dependence of cleavage rate on DNA length~\cite{Jeltsch96}, but from the same enzyme's processivity a much shorter sliding length of about 50 bp was estimated later~\cite{Stanford00}. The DNA concentration in the latter work was 5~nM for a 690~bp DNA, while the highest chain density used here was 0.4~nM for $L = 56$ persistence lengths, corresponding to an 8230~bp DNA. For the DNA length and concentration used in~\cite{Stanford00}, $\xi_{\text{opt}} = 0.22$, or 33 bp. We thus see that the relatively short sliding lengths estimated in more recent work make good sense for the biological function of DNA-binding proteins, since they constitute the best compromise between one- and three-dimensional search. 

The limits of our new approach are presently under investigation.
In the derivation of Eq.~(\ref{eq30}) we assumed chemical equilibrium
between the free and the non-specifically bound states of the walker.
For high affinity of the protein to the DNA, i.e.\
large values of $\xi$, this assumption may not be justified,
since the protein always starts in free diffusion mode at
the periphery of the cell. The violation of that assumption
may become more serious if the chain density inside the
cell is low, so that the protein has to search for a long time
before it is able to bind to the DNA for the first time.
Additionally, in order to evaluate the efficiency of
1d-diffusion [Eq.~(\ref{eq25})], it was assumed that the
DNA axis could be considered as a straight line over
the distance of 1d-diffusion. This
is satisfied if the sliding length is smaller than
the persistence length of the chain, i.e., $\xi < 1$.

In summary, the relation (\ref{eq32}), derived from first principles,
provides a quantitative estimate for the reaction time of a protein that is moving 
under the control of two competitive transport mechanisms in a crowded environment. 
Although drawing an idealized picture of the living cell, it will serve as the starting point
for more realistic approaches, equipped with additional parameters that are 
subsequently calibrated in sophisticated simulations. The sliding parameter 
$\xi$ [Eq.~(\ref{eq35})] connects the heuristic sliding
length of Halford et al.\ to experimentally accessible
quantities. The simulations, although so far performed on a
limited range of system parameters, confirm earlier results
that an unspecific binding combined with a 1d-diffusion mode
enables for a significant speed-up of the reaction.
The relation (\ref{eq32}) can be used to extend the
investigations to system sizes which are not
easily accessible in numerical simulations such as
those presented in this work: The size of
a realistic cell nucleus is of the order of ten microns
and it contains DNA chains adding up to a length of the
order of meters.

\begin{acknowledgments}
We thank J.\ F.\ Marko for fruitful discussions.
\end{acknowledgments}


\begin{thebibliography}{9}
\expandafter\ifx\csname natexlab\endcsname\relax\def\natexlab#1{#1}\fi
\expandafter\ifx\csname bibnamefont\endcsname\relax
  \def\bibnamefont#1{#1}\fi
\expandafter\ifx\csname bibfnamefont\endcsname\relax
  \def\bibfnamefont#1{#1}\fi
\expandafter\ifx\csname citenamefont\endcsname\relax
  \def\citenamefont#1{#1}\fi
\expandafter\ifx\csname url\endcsname\relax
  \def\url#1{\texttt{#1}}\fi
\expandafter\ifx\csname urlprefix\endcsname\relax\def\urlprefix{URL }\fi
\providecommand{\bibinfo}[2]{#2}
\providecommand{\eprint}[2][]{\url{#2}}

\bibitem[{\citenamefont{Riggs et~al.}(1970)\citenamefont{Riggs, Bourgeois, and
  Cohn}}]{riggs70}
\bibinfo{author}{\bibfnamefont{A.~D.} \bibnamefont{Riggs}},
  \bibinfo{author}{\bibfnamefont{S.}~\bibnamefont{Bourgeois}},
  \bibnamefont{and} \bibinfo{author}{\bibfnamefont{M.}~\bibnamefont{Cohn}},
  \bibinfo{journal}{J.\ Mol.\ Biol.} \textbf{\bibinfo{volume}{53}},
  \bibinfo{pages}{401} (\bibinfo{year}{1970}).

\bibitem[{\citenamefont{Richter and Eigen}(1974)}]{richter74}
\bibinfo{author}{\bibfnamefont{P.~H.} \bibnamefont{Richter}} \bibnamefont{and}
  \bibinfo{author}{\bibfnamefont{M.}~\bibnamefont{Eigen}},
  \bibinfo{journal}{Biophys.\ Chem.} \textbf{\bibinfo{volume}{2}},
  \bibinfo{pages}{255} (\bibinfo{year}{1974}).

\bibitem[{\citenamefont{Berg and von Hippel}(1985)}]{berg85}
\bibinfo{author}{\bibfnamefont{O.~G.} \bibnamefont{Berg}} \bibnamefont{and}
  \bibinfo{author}{\bibfnamefont{P.~H.} \bibnamefont{von Hippel}},
  \bibinfo{journal}{Annu.\ Rev.\ Biophys.\ Chem.}
  \textbf{\bibinfo{volume}{14}}, \bibinfo{pages}{130} (\bibinfo{year}{1985}).

\bibitem[{\citenamefont{Ptashne and Gann}(2001)}]{ptashne01}
\bibinfo{author}{\bibfnamefont{M.}~\bibnamefont{Ptashne}} \bibnamefont{and}
  \bibinfo{author}{\bibfnamefont{A.}~\bibnamefont{Gann}},
  \emph{\bibinfo{title}{Genes and Signals}} (\bibinfo{publisher}{Cold Spring
  Harbor Laboratory Press}, \bibinfo{address}{Cold Spring Harbor, NY},
  \bibinfo{year}{2001}).

\bibitem[{\citenamefont{Berg et~al.}(1981)\citenamefont{Berg, Winter, and von
  Hippel}}]{berg81}
\bibinfo{author}{\bibfnamefont{O.~G.} \bibnamefont{Berg}},
  \bibinfo{author}{\bibfnamefont{R.~B.} \bibnamefont{Winter}},
  \bibnamefont{and} \bibinfo{author}{\bibfnamefont{P.~H.} \bibnamefont{von
  Hippel}}, \bibinfo{journal}{Biochemistry} \textbf{\bibinfo{volume}{20}},
  \bibinfo{pages}{6929} (\bibinfo{year}{1981}).

\bibitem[{\citenamefont{Halford and Marko}(2004)}]{halford04}
\bibinfo{author}{\bibfnamefont{S.~E.} \bibnamefont{Halford}} \bibnamefont{and}
  \bibinfo{author}{\bibfnamefont{J.~F.} \bibnamefont{Marko}},
  \bibinfo{journal}{Nucl. Acids Res.} \textbf{\bibinfo{volume}{32}},
  \bibinfo{pages}{3040} (\bibinfo{year}{2004}).

\bibitem[{\citenamefont{Szabo et~al.}(1980)\citenamefont{Szabo, Schulten, and
  Schulten}}]{szabo80}
\bibinfo{author}{\bibfnamefont{A.}~\bibnamefont{Szabo}},
  \bibinfo{author}{\bibfnamefont{K.}~\bibnamefont{Schulten}}, \bibnamefont{and}
  \bibinfo{author}{\bibfnamefont{Z.}~\bibnamefont{Schulten}},
  \bibinfo{journal}{J.\ Chem.\ Phys.} \textbf{\bibinfo{volume}{72}},
  \bibinfo{pages}{4350} (\bibinfo{year}{1980}).

\bibitem[{\citenamefont{Klenin and Langowski}(2004)}]{klenin04}
\bibinfo{author}{\bibfnamefont{K.~V.} \bibnamefont{Klenin}} \bibnamefont{and}
  \bibinfo{author}{\bibfnamefont{J.}~\bibnamefont{Langowski}},
  \bibinfo{journal}{J.\ Chem.\ Phys.} \textbf{\bibinfo{volume}{121}},
  \bibinfo{pages}{4951} (\bibinfo{year}{2004}).

\bibitem[{\citenamefont{Klenin et~al.}(1998)\citenamefont{Klenin, Merlitz, and
  Langowski}}]{klenin98}
\bibinfo{author}{\bibfnamefont{K.}~\bibnamefont{Klenin}},
  \bibinfo{author}{\bibfnamefont{H.}~\bibnamefont{Merlitz}}, \bibnamefont{and}
  \bibinfo{author}{\bibfnamefont{J.}~\bibnamefont{Langowski}},
  \bibinfo{journal}{Biophys.\ J.} \textbf{\bibinfo{volume}{74}},
  \bibinfo{pages}{780} (\bibinfo{year}{1998}).

\bibitem[{\citenamefont{Jeltsch et~al.}(1998)\citenamefont{Jeltsch, Wenz, Stahl and Pingoud}}]{Jeltsch96}
\bibinfo{author}{\bibfnamefont{A.}~\bibnamefont{Jeltsch}},
  \bibinfo{author}{\bibfnamefont{C.}~\bibnamefont{Wenz}}, 
  \bibinfo{author}{\bibfnamefont{F.}~\bibnamefont{Stahl}},   \bibnamefont{and}
  \bibinfo{author}{\bibfnamefont{A.}~\bibnamefont{Pingoud}},
  \bibinfo{journal}{EMBO~J.} \textbf{\bibinfo{volume}{15}},
  \bibinfo{pages}{5104} (\bibinfo{year}{1996}).

\bibitem[{\citenamefont{Stanford et~al.}(1998)\citenamefont{Stanford, Szczelkun, Marko and Halford}}]{Stanford00}
\bibinfo{author}{\bibfnamefont{N.~P.}~\bibnamefont{Stanford}},
  \bibinfo{author}{\bibfnamefont{M.~D.}~\bibnamefont{Szczelkun}}, 
  \bibinfo{author}{\bibfnamefont{J.~F.}~\bibnamefont{Marko}},   \bibnamefont{and}
  \bibinfo{author}{\bibfnamefont{S.~E.}~\bibnamefont{Halford}},
  \bibinfo{journal}{EMBO~J.} \textbf{\bibinfo{volume}{19}},
  \bibinfo{pages}{6546} (\bibinfo{year}{2000}).

\end{thebibliography}

\end{document}